\documentclass[12pt]{amsart}
\usepackage[utf8]{inputenc}
\usepackage{xcolor}

\usepackage{geometry}                		
\theoremstyle{definition}
\newtheorem{dfn}{Definition}[section]
\newtheorem{rem}[dfn]{Remark}

\theoremstyle{plain}
\newtheorem{thm}[dfn]{Theorem}
\newtheorem{prop}[dfn]{Proposition}

\begin{document}

\title[Berezin quantization, conformal welding and Bott-Virasoro group]{Berezin quantization, conformal welding and the Bott-Virasoro group }

\author{A. Alekseev}
\address{Section of Mathematics, University of Geneva, Rue du Conseil G\'en\'eral 7-9, 12211, Geneva, Swtizerland}
\email{Anton.Alekseev@unige.ch}

\author{S. Shatashvili}
\address{The Hamilton Mathematics Institute, Trinity College Dublin, Dublin 2, Ireland}
\address{The School of Mathematics, Trinity College Dublin, Dublin 2, Ireland}
\address{Simons Center for Geometry and Physics, Stony Brook, USA}
\email{samson@math.tcd.ie}

\author{L. Takhtajan}
\address{Department of Mathematics, Stony Brook University, NY 11794-3651, USA}
\address{
Euler International Mathematical Institute, Pesochnaya Nab. 10, Saint Petersburg 197022, Russia}
\email{leontak@math.stonybrook.edu}


\begin{abstract}
Following Nag-Sullivan, we study the representation of the group ${\rm Diff}^+(S^1)$ of diffeomorphisms of the circle on the Hilbert space of holomorphic functions. Conformal welding provides a triangular decompositions for the corresponding symplectic transformations. We apply Berezin formalism and lift this decomposition to operators acting on the Fock space. This lift provides quantization of conformal welding, gives a new representative of the Bott-Virasoso cocylce class, and leads to a surprising identity for the Takhtajan-Teo energy functional on ${\rm Diff}^+(S^1)$. 
\end{abstract}

\dedicatory{Dedicated to the memory of our old friend Krzysztof Gawedzki}


\maketitle

\section{Introduction}

Coadjoint orbits of the canonical central extension 
$$
1 \to S^1 \to \widehat{{\rm Diff}^+(S^1)} \to {\rm Diff}^+(S^1) \to 1
$$
of the group $\mathcal{G}={\rm Diff}^+(S^1)$ of orientation preserving diffeomorphisms of the circle (also called Virasoro coadjoint orbits) attracted attention both in mathematics and physics literature since long time, see {\em e.g.} \cite{AS1, Kirillov, LP, Segal, W1}. The coadjoint action on the hyperplane corresponding to the coordinate $c$ (dual to 
${\rm Lie}(S^1) \cong \mathbb{R}$) is defined on the space of quadratic differentials on the circle $T(x) dx^2$, and it is given by formula
$$
\chi: T(x) dx^2 \mapsto T^\chi(x) dx^2 = \left(T(\chi(x)) \chi'(x)^2 + \frac{c}{12} \, {\rm Sch}(\chi)\right)  dx^2,
$$
where ${\rm Sch}(\chi)$ is the Schwarzian derivative
$$
{\rm Sch}(\chi) = \frac{\chi'''(x)}{\chi'(x)} - \frac{3}{2} \left( \frac{\chi''(x)}{\chi'(x)} \right)^2.
$$

For $c\neq 0$, one of the Virasoro coadjoint orbits is of special importance. It corresponds to $T(x)=\frac{c}{24} \, dx^2$, and it is the unique orbit with the stabiliser isomorphic to the group ${\rm PSL}(2, \mathbb{R})$. This orbit (also called the Teichm\"uller orbit) naturally embeds in the universal Teichm\"uller space $T(1)$\footnote{Teichm\"uller spaces for curves of all finite genera naturally embed in $T(1)$.}:
$$
O_{\rm Teich}  \cong {\rm Diff}^+(S^1)/{\rm PSL}(2, \mathbb{R}) \subset {\rm QS}(S^1)/{\rm PSL}(2, \mathbb{R}),
$$
where ${\rm QS}(S^1)$ is the group of quasi-conformal mappings of the circle.

Consider the space ${\rm Hyp}(\mathbb{D})$ of geodesically complete hyperbolic metrics on the unit disk $\mathbb{D} \subset \mathbb{C}$. A typical example in this class is the standard Poincar\'e metric.
The group of orientation preserving diffeomorphisms ${\rm Diff}^+(\mathbb{D})$ acts transitively on ${\rm Hyp}(\mathbb{D})$. Now consider the group ${\rm Diff}^+(\mathbb{D}, \partial \mathbb{D}) \subset {\rm Diff}^+(\mathbb{D})$ which fixes the boundary of the disk 
$\partial \mathbb{D} \cong S^1$. 
It was argued in the physics literature (see \cite{SSS}) that $O_{\rm Teich}$ is  symplectomorphic to the following moduli space:
$$
O_{\rm Teich} \cong {\rm Hyp}(\mathbb{D})/{\rm Diff}^+(\mathbb{D}, \partial \mathbb{D}).
$$
Formal Duistermaat-Heckman integrals over this space were defined and studied in \cite{Stanford_Witten, Alekseev_Shatashvili_char_orbits_DH}.


Recall that for $\chi \in {\rm QS}(S^1)$ there exist two univalent holomorphic functions
$f_+: \mathbb{D} \to \mathbb{C}, f_-: \mathbb{D}^* \to \mathbb{C}$ such that
$$
f_+(e^{i \chi(x)}) = f_-(e^{ix})
$$
for $x \in \mathbb{R}$. Here $\mathbb{D}^*$ is the unit disk centered at infinity. The functions $f_+(z)$ and $f_-(z)$ are called components of conformal welding of $\chi$. 
The K\"ahler potential of the Weil-Petersson metric on $O_{\rm Teich}$ is given by the Takhtajan-Teo (TT) energy functional (see \cite{TT}\footnote{In fact, in \cite{TT} a new Hilbert manifold structure on $T(1)$ was introduced, and it was shown that $S(\chi)$ is a K\"ahler potential for the Weil-Petersson metric on the Hilbert submanifold $T_{0}(1) \subset T(1)$.}):
$$
S(\chi) = \int_\mathbb{D} \left| \frac{f''_+(z)}{f'_+(z)} \right|^2 \, d^2z + 
4 \pi \log(|f'_+(0)|) + 
\int_{\bar{\mathbb{D}}} \left| \frac{f''_-(z)}{f'_-(z)} \right|^2 \, d^2z - 
4 \pi \log(|f'_-(\infty)|).
$$

In this paper, we focus our attention on the subgroup ${\rm Diff}^+_{\rm hol}(S^1) \subset 
{\rm Diff}^+(S^1)$ which is characterized by the property that the map $z=e^{ix} \mapsto e^{i\chi(x)}$ extends to a holomorphic function on an annulus $\mathcal{A}_{r,R}=\{ z \in \mathbb{C}; r < |z| < R\}$ for some $r <1 <R$.
 For this subgroup, following Nag-Sullivan \cite{Sullivan} we define a group homomorphism to the group of restricted symplectic transformations acting on the Hilbert space $H=H_+ \oplus H_-$ of holomorphic functions (modulo constants):
$$
{\rm Diff}^+_{\rm hol}(S^1) \to {\rm Sp}^{\rm res}(H_+ \oplus H_-).
$$
Here $H_+$ is spanned by $z^n$ for $n\geq 1$, and $H_-$ by $z^n$ for $n \leq -1$, and $||z^n||=|n|$.
We then use the metaplectic representation of ${\rm Sp}^{\rm res}(H_+ \oplus H_-)$ defined by Berezin formalism of normal symbols (see \cite{Berezin}) to construct operators
\begin{equation}  \label{eq:quantum_welding}
N_\chi = N_{f_+^{-1}} * N_{f_-}.
\end{equation}
Here $*$ is the product of operators acting on the Fock space $\mathcal{F}$ defined by the polarization $H = H_+ \oplus H_-$.  In this sense, equation \eqref{eq:quantum_welding} defines a  quantization of conformal welding. 

Our first main result is as follows:
\begin{thm}  \label{thm:intro1}
For $\chi, \phi \in {\rm Diff}^+_{\rm hol}(S^1)$, we have
$$
N_\chi N_\phi = C(\chi, \phi) N_{\chi \circ  \phi},
$$
where $C(\chi, \phi) \in \mathbb{C}^*$ is a multiplicative group 2-cocylce with 
the property that
\begin{equation}  \label{eq:intro_surprize}
C(\chi, \phi) = C(f_-, g_+),
\end{equation}
and $f_\pm$ define a conformal welding of $\chi$, and $g_\pm$ define a conformal welding of $\phi$.
Furthermore, for $\chi \in {\rm Diff}^+_{\rm hol}(S^1)$
$$
\log(|C(f_-^{-1}, f_+)|) = \frac{S(\chi)}{24\pi}.
$$
\end{thm}
Equation \eqref{eq:intro_surprize} is surprizing since the cocycle $C(\chi, \phi)$ depends only on the components $f_-$ and $g_+$! For a more precise statement of Theorem \ref{thm:intro1}, see Theorem \ref{thm:C_N}
and Theorem \ref{thm:C_N_welding}.

Our second main result is the following theorem (see also Theorem \ref{thm:main2}):
\begin{thm}  \label{thm:intro2}
For $\chi \in {\rm Diff}^+_{\rm hol}(S^1)$, the operator
$$
U_\chi = e^{S(\chi)/48\pi} N_\chi
$$
is unitary, and the cocycle $C(\chi, \phi)$ satisfies the equality
\begin{equation}    \label{eq:mystery}
\log(|C(\chi, \phi)|) = \frac{S(\chi \circ \phi) - S(\chi) - S(\phi)}{48\pi}.
\end{equation}
\end{thm}
The left hand side and the right hand side of \eqref{eq:mystery} have rather different analytic forms.  Equation \eqref{eq:mystery} follows from Berezin formalism, but at this point we are not aware of its direct proof.

We believe that our findings admit extensions to other representations of the group of diffeomorphisms of the circle defined in terms of free fields. In particular, this applies to representations of affine Kac-Moody algebras on Wakimoto modules. We also believe that our results may find applications in Theoretical Physics. Our original motivation comes from the work \cite{Polyakov} which introduced conformal welding in the study of Fermions in a gravitational field in 2 dimensions (see also the analysis of the gravitational Wess-Zumino functionals in \cite{AS2}).

The structure of the paper is as follows: in Section 2, we recall the definition of the Bott-Virasoro 2-cocycle on ${\rm Diff}^+(S^1)$, and we extend it to the groupoid of conformal maps.
In Section 3, we explain how holomorphic maps define symplectic transformations on the space $H$, and we define their Grunsky coefficients. In Section 4, we set up the Berezin formalism for normal and unitary symbols of operators on the Fock space $\mathcal{F}$. Finally, in Section 5 we describe quantization of conformal welding, the cocycles for normal and unitary symbols and their relation to the TT functional.

\vskip 0.2cm

{\bf Acknowledgements:} 
Research of AA was supported in part by the grants 182767, 208235 and 200400, 
and by the NCCR SwissMAP of the Swiss National Science Foundation (SNSF).
Research of AA and SS was partly supported by the award of
the Simons Foundation to the Hamilton Mathematics Institute of the Trinity College Dublin
under the program ``Targeted Grants to Institutes''.

\section{Group cocycles}

In this Section, we recall the notion of a group 2-cocycle. We then focus our attention on the Bott-Virasoro cocycle on the group of orientation preserving diffeomorphisms of the circle ${\rm Diff}^+(S^1)$ and on its extension to holomorphic maps.

\subsection{Group 2-cocycles: definition and basic properties}

Let $G$ be a group and $\mathbb{K}$ be the basic field (in this article, $\mathbb{R}$ or $\mathbb{C}$) viewed as a trivial $G$-module. A map $c: G \times G  \to \mathbb{K}$ is an additive group 2-cocycle if
\begin{equation} \label{eq:2-cocycle}
c(f,g) + c(fg, h) = c(f, gh) + c(g,h)
\end{equation}
for all $f, g, h \in G$. The definition implies $c(e,g)=c(e,e)=c(g,e)$ for all $g \in G$. Also, for all $k \in \mathbb{K}$ the assignment $c(f,g)=k$ is a 2-cocycle.

For every map $b: G \to \mathbb{K}$ one defines a trivial 2-cocycle
$$
\delta b(f, g) = b(f) - b(fg) + b(g).
$$
Note that for $b(f) =k \in \mathbb{K}$ we obtain $c(f,g)=k$. Hence, for any 2-cocycle $c$ there is a cohomologous normalized cocycle
$$
\tilde{c}(f,g) = c(f,g) - c(e,e)
$$
which has the property $\tilde{c}(e,e)=0$.
We will use the following cyclic property of 2-cocycles:
\begin{prop}  \label{prop:cyclic}
Assume that a 2-cocycle $c$ has the property $c(f, f^{-1})=0$ for all $f \in G$. Then,
\begin{equation}  \label{eq:cyclic}
c(f, g) = c(g, h) = c(h, f),
\end{equation}
where $fgh=e$.
\end{prop}

\begin{proof}
In equation \eqref{eq:2-cocycle}, put $fgh=e$ to obtain
$$
c(f, g) + c(h^{-1}, h) = c(f, f^{-1}) + c(g, h).
$$
By assumption, $c(h^{-1}, h)=c(f, f^{-1})=0$. Hence, we get $c(f, g)=c(g, h)$. The last equality follows since 
$fgh=e$ implies $hfg=e$.
\end{proof}

Assume that the group $G$ possesses a Lie algebra $\mathfrak{g} = {\rm Lie}(G)$, denote by $\exp: \mathfrak{g} \to G$ the exponential map, and assume that finite products of the type
$\exp(u_1) \dots \exp(u_m)$ cover $G$. We define a map $\beta: \mathfrak{g} \times G \to \mathbb{K}$ by formula
\begin{equation}
\beta(u, g)=\frac{d}{dt} \, c(\exp(tu), g)|_{t=0}.
\end{equation}
The condition $c(g, e)=c(e,e)$ implies that $\beta(u, e)=0$ for all $u \in \mathfrak{g}$.

\begin{prop} \label{prop:b=>c}
The map $\beta: \mathfrak{g} \times G \to \mathbb{K}$ uniquely determines a normalized group 2-cocycle $c$.
\end{prop}

\begin{proof}
Put $f=\exp(tu)$ in equation \eqref{eq:2-cocycle} and differentiate in $t$ at $t=0$. We obtain
$$
\frac{d}{dt} c(\exp(tu)g, h) = \beta(u, gh) - \beta(u,g).
$$
Hence, if $\beta(u,g)=0$ then the normalized cocycle $c(f,g)$ vanishes, as required.
\end{proof}

Define 
$$
\alpha(u,v) = \frac{d}{ds} \, \beta(u, \exp(sv))|_{s=0}
=\frac{\partial^2}{\partial s \partial t} \, c(\exp(tu), \exp(sv))|_{s=t=0}
$$
and 
$$
a(u,v) = \frac{1}{2}\left(\alpha(u,v) - \alpha(v,u)\right).
$$
Recall that $a \in \wedge^2 \mathfrak{g}^*$ is a Lie algebra 2-cocycle, and that it satisfies the equation
\begin{equation}
a(u, [v, w]) + a(w, [u, v]) + a(v, [w, u])=0.    
\end{equation}
Note that there is no analog of Proposition \ref{prop:b=>c} which would allow to reconstruct maps $\beta$ and $c$ starting from the map $a$ (or the map $\alpha$). Indeed,
adding a trivial cocycle $\delta b$ such that $d/dt \, b(\exp(tu))|_{t=0} =0$ does not affect the maps $\alpha$ and $a$, but it changes $\beta$ and $c$, in general.

Let $\rho: G \to {\rm End}(V)$ be a  projective representation, and assume that
$$
\rho(f) \rho(g) = e^{i c(f, g)} \rho(fg),
$$
where $c: G \times G \to \mathbb{C}$ is a complex valued function. Then, $c$ verifies the identity \eqref{eq:2-cocycle} modulo $2\pi \mathbb{Z}$. This is a direct consequence of associativity of the product in ${\rm End}(V)$.
If the group $G$ is connected, and the function $c$ is a continuous function, then it is actually a 2-cocycle. Indeed, in this case the defect in equation \eqref{eq:2-cocycle} is also a continuous function of $f, g, h$ which vanishes for $f=g=h=e$. Hence, it vanishes for all $f, g, h \in G$.
Furthermore, assume that $V$ is a Hilbert space and that $\rho: G \to U(V)$ is a unitary representation. Then, $c(f,g) \in \mathbb{R}$ is a real valued 2-cocycle.

Every 2-cocycle $c: G \times G \to \mathbb{C}$ defines a group law on $\hat{G} = G \times \mathbb{C}^*$ defined by formula
$$
(f, z) \cdot (g, w) = (fg, zw \exp(ic(f,g))).
$$
This group fits into a short exact sequence
$$
1 \to \mathbb{C}^* \to \hat{G} \to G \to 1
$$
and defines a central extension of $G$. If the cocycle $c$ is real valued, this central extension is by the circle $S^1$ (instead of $\mathbb{C}^*$).

\subsection{The Bott-Virasoro cocycle}

Consider the group $G={\rm Diff}^+(S^1)$ of orientation preserving diffeomorphisms of the circle. We recall the following basic fact:
\begin{thm}[Bott-Virasoro cocycle]  \label{thm:Bott-Virasoro}
The map $c_{\rm BV}: G \times G \to \mathbb{R}$ defined by formula
$$
c_{\rm BV}(\chi,\phi) = \int_0^{2\pi} \log(\chi'(\phi(x))) \log(\phi'(x))'\,  dx
$$
is a normalized real valued group 2-cocycle. 
Furthermore, it satisfies the cyclic property \eqref{eq:cyclic}.
\end{thm}

\begin{proof} 
For convenience of the reader, we give a proof of this statement.
The left hand side of equation \eqref{eq:2-cocycle} is as follows:
$$
\begin{array}{lll}
 c_{\rm BV}(\chi,\phi)+c_{\rm BV}(\chi \circ \phi,\psi) & =    &
 \int_0^{2\pi}\left( \log(\chi'(\phi(x))(\log(\phi'(x)))' + \log((\chi\circ \phi)'(\psi(x))) (\log(\psi'(x)))' \right) \, dx \\
  & = & \int_0^{2\pi} \left( \log(\chi'(\phi(\psi(x)))) \log(\psi'(x))' + \log(\phi'(\psi(x))) \log(\psi'(x))'\right) \, dx \\
  & +  & \int_0^{2\pi} \log(\chi'(\phi(x))(\log(\phi'(x)))'\, dx, \\
\end{array}
$$
and the right hand has the following form:
$$
\begin{array}{lll}
c_{\rm BV}(\chi, \phi \circ \psi) + c_{\rm BV}(\phi,\psi)  & = &
\int_0^{2\pi}\left( \log(\chi'(\phi(\psi(x)))) \log((\phi\circ \psi)'(x))' +
\log(\phi'(\psi(x))) \log(\psi'(x))'\right) \, dx\\
& = & \int_0^{2\pi}\left( \log(\chi'(\phi(\psi(x)))) \log(\psi'(x))' + \log(\phi'(\psi(x))) \log(\psi'(x))'\right) \, dx \\
& + & \int_0^{2\pi} \log(\chi'(\phi(\psi(x))) \log(\phi'(\psi(x)))' \, dx.
\end{array}
$$
Note that the second lines in the two expressions coincide term by term, and the third line of the right hand side of \eqref{eq:2-cocycle} is obtained from the third line of the left hand side by the change of variable $x \mapsto \psi(x)$.

For the cyclic property, put $\chi = \phi^{-1}$. Then, $\log(\chi'(\phi(x))) = - \log(\phi'(x))$ and
$$
c_{\rm BV}(\phi^{-1}, \phi) = - \int_0^{2\pi} \log(\phi'(x)) \log(\phi'(x))' \, dx = - \frac{1}{2} \, \log(\phi'(x))^2|_0^{2\pi} = 0.
$$
Here we have used the fact that $\phi'(x)$ is periodic. The cyclic property follows by Proposition \ref{prop:cyclic}. 
\end{proof}

It is instructive to compute the map $\beta_{\rm BV}$:
$$
\beta_{\rm BV}(u,\phi) = \int_0^{2\pi} u'(\phi(x)) \log(\phi'(x))' \, dx =
- \int_0^{2\pi} u'(y) \log((\phi^{-1})'(y))' \, dy.
$$
Here we made a change of variables $y=\phi(x)$ in the integral.
The map $\alpha_{\rm BV}$ is given by formula,
$$
\alpha_{\rm BV}(u,v) = \int_0^{2\pi} u'(x) v''(x) dx,
$$
and it is skew-symmetric in $u$ and $v$.

\subsection{Extension to holomorphic maps}
In this Section, we extend the Bott-Virasoro cocycle to a certain class of holomorphic maps. In more detail, let $\mathcal{S}$ be a connected open subset of the complex plane with $\pi_1(\mathcal{S}) \cong \mathbb{Z}$. By the Uniformization Theorem, such a domain is holomorphically isomorphic to an annulus
$$
\mathcal{A}_{r, R} = \{ z \in \mathbb{C}; r < |z| <R \} .
$$
We will consider triples $(\mathcal{S}, f, \mathcal{T})$, where $\mathcal{S}$ and $\mathcal{T}$ are two such domains, and $f: \mathcal{S} \to \mathcal{T}$ is a holomorphic isomorphism between $\mathcal{S}$ and $\mathcal{T}$. Sometimes it is convenient to label the domain and the range of $f$ by $\mathcal{S}_f$ and $\mathcal{T}_f$, respectively. Note that the domain $\mathcal{S}_f$ contains a closed curve $C_f$ which represents the generator of $\pi_{1}(\mathcal{S}_f)$. Its image $f(C_f)$ represents the generator of $\pi_1(\mathcal{T}_f) \cong \mathbb{Z}$. By the analytic continuation principle, the holomorphic function $f$ is uniquely determined by its restriction to $C_f$.

Triples $(\mathcal{S}_f, f, \mathcal{T}_f)$ form a groupoid with composition law
$$
(\mathcal{S}_f, f, \mathcal{T}_f) \circ (\mathcal{S}_g, g, \mathcal{T}_g) =
(\mathcal{S}_g, f\circ g, \mathcal{T}_f).
$$
Two triples are composable if $\mathcal{S}_f = \mathcal{T}_g$. The curves $g(C_g)$ and $C_f$ are homotopic to each other in $\mathcal{S}_f = \mathcal{T}_g$.

We define a subset ${\rm Diff}^+_{\rm hol}(S^1) \subset {\rm Diff}^+(S^1)$ by the following property: $\chi \in {\rm Diff}^+_{\rm hol}(S^1)$ if the map
$$
z = e^{ix} \mapsto e^{i \chi(x)}
$$
extends to a univalent holomorphic map $f$ on an annulus $\mathcal{A}_{r, R}$ with $r < 1 < R$. It is easy to see that ${\rm Diff}^+_{\rm hol}(S^1)$ is a subgroup of ${\rm Diff}^+(S^1)$.

%
%

Consider two diffeomorphisms $\chi, \phi \in {\rm Diff}^+_{\rm hol}(S^1)$ and the corresponding univalent holomorphic functions $f$ and $g$ such that
$$
f(e^{ix})=e^{i\chi(x)}, \hskip 0.3cm g(e^{ix}) = e^{i \phi(x)}.
$$
By making the annulus $\mathcal{S}_g$ smaller if needed, one can always achieve $\mathcal{T}_g \subset \mathcal{S}_f$. By restricting $f$ to $\mathcal{T}_g$, one obtains a pair of composable holomorphic maps, and
$$
f(g(e^{ix})) = f(e^{i\phi(x)}) = e^{i\chi(\phi(x))}.
$$
Since the analytic function $f \circ g$ is uniquely determined by its values on the unit circle, we conclude that it corresponds to the diffeomorphism $\chi \circ \phi$.

\begin{prop}  \label{prop:cBVextend}
Let $\chi, \phi \in {\rm Diff}^+_{\rm hol}(S^1)$ and $f,g$ the corresponding composable holomorphic functions. Then,
\begin{equation}  \label{eq:cBVextend}
c_{\rm BV}(\chi, \phi) = \int_{C_1} \log\left( \frac{g(z)f'(g(z))}{f(g(z))} \right) \,
\log\left( \frac{z g'(z)}{g(z)}\right)' \, dz.
\end{equation}
\end{prop}

\begin{proof}
The proof is by a direct calculation. In particular, for $z=e^{ix}, g(z)=e^{i\phi(x)}$ we have $zg'(z)/g(z)=\phi'(x)$, and  $g(z)f'(g(z))/f(g(z)) = \chi'(\phi(x))$.
\end{proof}

For a pair of composable univalent holomorphic maps $f$ and $g$, one can use the right hand side of equation \eqref{eq:cBVextend} as a definition of a  functional of a pair $(f,g)$:
\begin{equation}  \label{eq:CBV}
C_{\rm BV}(f,g)=\int_{C_g} \log\left( \frac{g(z)f'(g(z))}{f(g(z))} \right) \,
\log\left( \frac{z g'(z)}{g(z)}\right)' \, dz.
\end{equation}
Here the integration is over the curve $C_g$ on which both homorphic functions $g$ and $f \circ g$ are well defined.
Note that $C_{\rm BV}$ is complex valued, in general. This is in contrast to the cocycle $c_{\rm BV}$ which takes values in $\mathbb{R}$. 

\begin{prop}  \label{prop:C_BV}
The map $C_{\rm BV}$ is a groupoid 2-cocylce. That is, for all composable triples $(f,g,h)$, it satisfies the equation
$$
C_{\rm BV}(f,g)+C_{\rm BV}(fg, h) = C_{\rm BV}(f, gh) + C_{\rm BV}(g,h).
$$
Furthermore, it satisfies the cyclic property \eqref{eq:cyclic}.
\end{prop}

\begin{proof}
The proof of the cocycle condition is analogous to the one of Theorem \ref{thm:Bott-Virasoro}. The only non-trivial step in the proof is as follows: one needs to check that
$$
\int_{C_h} \log\left( \frac{g(h(z))f'(g(h(z)))}{f(g(h(z)))} \right) \,
\log\left(\frac{h(z)g'(h(z))}{g(h(z))}\right)' \, dz =
\int_{C_g} \log\left( \frac{g(w)f'(g(w))}{f(g(w))}\right) \, \log\left( \frac{wg'(w)}{g(w)} \right)' dw.
$$
Two integrals are related by the change of variable $w=h(z)$. After this change of variables, the integration contour on the left hand side is $C_h$, and on the right hand side it is $h^{-1}(C_g)$. Both these curves represent the generator of $\pi_1(\mathcal{S}_h)$, and therefore they are homotopic to each other. 

For the cyclic property, put $f=g^{-1}$. Then, 
$$
\log\left( \frac{g(z)f'(g(z))}{f(g(z))} \right) = - \log\left( \frac{z g'(z)}{g(z)}\right)
$$
and 
$$
C_{\rm BV}(g^{-1}, g) = - \frac{1}{2} \int_{C_g} \frac{d}{dz} \, \left(\log\left( \frac{z g'(z)}{g(z)}\right)\right)^2 \, dz =0.
$$
The proof of Proposition \ref{prop:cyclic} applies verbatim to the case of groupoids. This completes the proof.
\end{proof}

Note that the expression \eqref{eq:CBV} can be re-written using  the change of variables $z=g^{-1}(w)$.  We get
$$
C_{\rm BV}(f, h^{-1}) = 
- \int_{C_f} \log\left( \frac{w f'(w)}{f(w)}\right) \, \log\left(
\frac{w (g^{-1})'(w)}{g^{-1}(w)}\right)' \, dw.
$$
Holomorphic functions $f: \mathcal{S}_f \to \mathcal{T}_f$ and $h: \mathcal{T}_g \to \mathcal{S}_g$ are actually defined on the same domain 
$\mathcal{S}_f = \mathcal{T}_g$ which contains the curve $C_f$. We compute the expression $\beta$ for $C_{\rm BV}$.
By putting $f(w)=w + t u(w) + O(t^2)$, we obtain
$$
\beta_{\rm BV}(u, h^{-1}) =
- \int_{C_1} \left(u'(w) - \frac{u(w)}{w}\right) \, \log\left(
\frac{w h'(w)}{h(w)}\right)' \, dw.
$$

%
%
%

\begin{rem}
Yet another groupoid cocycle which has the cyclic property is given by formula
\begin{equation}  \label{eq:tildeC}
\tilde{C}(f,g) = \int_{C_g} \log(f'(g(z))) \log(g'(z))' dz.
\end{equation}
The proof is similar to those of Theorem \ref{thm:Bott-Virasoro} and of Proposition \ref{prop:C_BV}.
\end{rem}

\section{Symplectic transformations}

In this Section, we recall the notion of symplectic transformations associated to holomorphic maps.

\subsection{Symplectic transformations in finite dimensions}

Recall the following standard setup: let $U$ be a complex vector space, and $\omega \in \wedge^2 U$ be a non-degenerate (symplectic) 2-form. A linear map $A \in {\rm End}(U)$ is called symplectic if it preserves $\omega$:
$$
\omega(A(u_1), A(u_2)) = \omega(u_1, u_2)
$$
for all $u_1, u_2 \in U$.

The following set of examples is of special interest for us: let $V$ be a finite dimensional complex vector space. Then, one can equip the direct sum
$U= V \oplus V^*$ with a  natural symplectic form
$$
\omega(a+a^*, b+b^*) = \langle b^* , a\rangle - \langle a^*, b\rangle,
$$
where $a,b \in V, a^*,b^* \in V^*$. A splitting $U= V \oplus V^*$ is also called a polarization of the symplectic space $U$.

Consider a transformation $A \in {\rm End}(V \oplus V^*)$ defined by formula
\begin{equation}      \label{eq:AinSp}
\left(
\begin{array}{ll}
a \\
a^*
\end{array}
\right) \mapsto
\left(
\begin{array}{ll}
\tilde{a} \\
\tilde{a}^*
\end{array}
\right) =
\left(
\begin{array}{ll}
\alpha & \beta \\
\gamma & \delta
\end{array}
\right) \cdot 
\left(
\begin{array}{ll}
a \\
a^*
\end{array}
\right).
\end{equation}
Here $\alpha: V \to V, \beta: V^*\to V, \gamma: V \to V^*, \delta: V^* \to V^*$.
This transformation is symplectic if and only if the following conditions are verified:
$$
\begin{array}{lll}
\beta \alpha^t = (\beta \alpha^t)^t, 
 &  \alpha^t \gamma = (\alpha^t \gamma)^t, 
& \alpha \delta^t - \beta \gamma^t = 1, \\
\delta^t \beta = (\delta^t \beta)^t, &
\gamma \delta^t = (\gamma \delta^t)^t, &
\alpha^t \delta - \gamma^t \beta = 1.
\end{array}
$$
Note that if $\alpha$ and $\delta$ are  invertible, the following operators are symmetric: $(\alpha^{-1} \beta)$, $(\beta \delta^{-1})$, $(\gamma \alpha^{-1})$, $(\delta^{-1} \gamma)$. In this case, one can also express $\alpha$ and  $\delta$ in terms of three other operators:
$$
\delta = (\alpha^t)^{-1} + \gamma \alpha^{-1} \beta, \hskip 0.3cm
\alpha = (\delta^t)^{-1} + \beta \delta^{-1} \gamma.
$$

\subsection{Symplectic transformations and holomorphic maps}
We now pass to the infinite dimensional context and apply the theory of symplectic transformations to holomorphic functions. As in the previous Section, let $\mathcal{S} \subset \mathbb{C}$ be a connected domain with $\pi_1(\mathcal{S})=\mathbb{Z}$, and let $C \subset \mathcal{S}$ be a closed oriented curve which realizes the generator of $\pi_1(\mathcal{S})$. We consider the space $\mathcal{H}_\mathcal{S}$ of holomorphic functions on $\mathcal{S}$ and define the following 2-form 
$$
\omega_\mathcal{S} = \frac{1}{4\pi} \int_C \, (\delta \phi(z)) \partial_z(\delta \phi(z)) \, dz.
$$
Here $\delta$ is the de Rham differential on $\mathcal{H}_\mathcal{S}$ and $\partial_z$ is the $z$-derivative. It is clear that the definition of $\omega_\mathcal{S}$ is independent of the choice of the curve $C$.

Let $f: \mathcal{S} \to \mathcal{T}$ be a holomorphic isomorphism. It induces an isomorphism $f^*: \mathcal{H}_\mathcal{T} \to \mathcal{H}_\mathcal{S}$ by composition: $\phi \mapsto f^*\phi(z) = \phi(f(z))$. In turn, the map $f^*$ induces a pull-back map of differential forms that we denote by $(f^*)^*$.

\begin{prop}
 $(f^*)^*\omega_\mathcal{S}=\omega_\mathcal{T}$.
\end{prop}

\begin{proof}
We compute,
$$
\begin{array}{lll}
4 \pi (f^*)^* \omega_\mathcal{S} & = &
\int_{C_\mathcal{S}} (\delta \phi(f(z))) \partial_z (\delta \phi(f(z))) \, dz \\
& = & 
\int_{f(C_\mathcal{S})} (\delta \phi(w)) \partial_w (\delta \phi(w)) \, dw \\
& = & 
\int_{C_\mathcal{T}} (\delta \phi(w)) \partial_w (\delta \phi(w)) \, dw \\
& = & 
4 \pi \omega_\mathcal{T}.
\end{array}
$$
Here we made a change of variables $z=f^{-1}(w)$, and then used the fact that 
$f(C_\mathcal{S})$ is homotopic of $C_\mathcal{T}$ in $\mathcal{T}$.
\end{proof}

For an annulus
$$
\mathcal{A}_{r, R} = \{ z \in \mathbb{C}; r < |z| < R\}
$$
with $r < 1 < R$, one can choose $C_\mathcal{S}$ to be the unit circle. 
We will consider the space $H = \mathcal{H}_\mathcal{A}/\mathbb{C}$ of holomorphic functions modulo constants.
Using the Fourier transform,
$$
\phi(z) = \sum_{n=1}^\infty \frac{a_n}{\sqrt{n}} z^n + \sum_{n=1}^\infty \frac{a_n^*}{\sqrt{n}}  z^{-n} + a_0,
$$
we obtain a formula for $\omega_\mathcal{A}$:
$$
\omega_\mathcal{A} =\frac{i}{2} \, \sum_{n =1}^\infty \delta a_n  \wedge \delta a_n^*.
$$
This form is symplectic on $H$. In what follows, it will be more convenient to work with functions
$$
\psi(z) = \phi'(z) =  \sum_{n=1}^\infty \sqrt{n} \, a_n z^{n-1} - \sum_{n=1}^\infty \sqrt{n} \, a^*_n z^{-n-1}
$$
which do not contain the superfluous constant $a_0$. One can view Fourier components $\{ a_n, a^*_n\}$ as coordinates on the infinite dimensional symplectic space of holomorphic functions.

The space $H$ admits a polarization
$$
H = H_+ \oplus H_-,
$$
where $H_+$ is spanned by monomials $z^n$ with $n \geq 1$ and $H_-$ by monomials $z^n$ with $n\leq -1$. Holomorphic maps induce symplectic transformations
$$
f: \psi \mapsto \tilde{\psi}(z)=\psi(f(z))f'(z).
$$
In more detail, 
\begin{equation}  \label{eq:f_acts_psi}
\sum_{n=1}^\infty \sqrt{n} \, (\tilde{a}_n z^{n-1} -
 \tilde{a}^*_n z^{-n-1})
 = f'(z) \sum_{n=1}^\infty \sqrt{n} ({a}_n (f(z))^{n-1} - {a}^*_n (f(z))^{-n-1}).
\end{equation}
This equation implies
\begin{equation} \label{eq:alpha}
\alpha_{m,n} = \frac{1}{2\pi i} \int_C \sqrt{\frac{n}{m}} 
\int_C \frac{f(z)^{n-1} f'(z)}{z^m} \, dz,
\end{equation}
\begin{equation} \label{eq:beta}
\beta_{m,n} = - \frac{1}{2\pi i} \int_C \sqrt{\frac{n}{m}} 
\int_C \frac{f(z)^{-n-1} f'(z)}{z^m} \, dz,
\end{equation}
\begin{equation} \label{eq:gamma}
\gamma_{m,n} = -\frac{1}{2\pi i} \int_C \sqrt{\frac{n}{m}} 
\int_C f(z)^{n-1} f'(z) z^m \, dz, 
\end{equation}
\begin{equation} \label{eq:delta}
\delta_{m,n} =  \frac{1}{2\pi i} \int_C \sqrt{\frac{n}{m}} 
\int_C f(z)^{-n-1} f'(z) z^m \, dz,
\end{equation}
where $\alpha_{m,n}, \beta_{m,n}, \gamma_{m,n}, \delta_{m,n}$ are infinite dimensional matrices representing operators $\alpha, \beta, \gamma, \delta$. 

The map from holomorphic maps to symplectic transformations is a group anti-homomorphism:
\begin{prop}  \label{prop:A_fg}
Let $f, g$ be two composable holomorphic maps and $A_f, A_g$ be the corresponding symplectic transformations. Then,
$$
A_{f \circ g} = A_g A_f.
$$
\end{prop}

\begin{proof}
The proof is by a direct computation.
\end{proof}

\subsection{Grunsky coefficients and symplectic transformations}

We will need the following simple properties of holomorphic functions.

Let $f(z)$ be a univalent holomorphic function on neighborhood of zero with $f(0)=0$. Then,  $f(z)=\sum_{n=1}^\infty f_n z^n$ with $f_1 \neq 0$. Recall that the function
\begin{equation}  \label{eq:Grunsky1}
\log\left(\frac{f(z)-f(w)}{z-w}\right) = \sum_{m,n=0}^\infty F_{m,n} z^m w^n
\end{equation}
is regular in $z$ and $w$. Here $F_{m,n}$ are the Grunsky coefficients of $f(z)$
(see \cite{Pom} for details).

In a similar fashion, let $f(z)$ be a univalent holomorphic function on a neighborhood of infinity with $f(\infty) = \infty$. Then, $f(z) = \sum_{n = - \infty}^1 f_n z^n$ with $f_1 \neq 0$, and
\begin{equation}   \label{eq:Grunsky2}
\log\left(\frac{f(z)-f(w)}{z-w}\right) = \log(f_1) + \sum_{m,n=1}^\infty F_{-m,-n} z^{-m} w^{-n}.
\end{equation}

%
%


%

%

The following proposition will be important for the rest of the paper:
\begin{prop}   \label{prop:f_h}
Let $f=\sum_{n=1}^\infty f_n z^n$ with $f_1 \neq 0$ be a univalent holomorphic map on a neighborhood of zero. Then, the corresponding symplectic transformation is upper-triangular, the operator $\gamma$ vanishes, and the symmetric operator $(\alpha^{-1} \beta)$ is of the following form:
\begin{equation}   \label{eq:TT_formula1}
\sum_{m,n=1}^\infty \sqrt{mn} \, (\alpha^{-1} \beta)_{m,n} u^{m-1} w^{-1}n = \frac{(f^{-1})'(u)(f^{-1})'(w)}{(f^{-1}(z)-f^{-1}(w))^2} - \frac{1}{(u-w)^2}.
\end{equation}
%

Let $f=\sum_{n=-\infty}^1 f_n z^n$ with $f_1 \neq 0$ be a univalent holomorphic map on a neighborhood of infinity. Then, the corresponding symplectic transformation is lower triangular, the operator $\beta$ vanishes, and the symmetric operator $(\gamma \alpha^{-1})$ is of the following form:
\begin{equation}   \label{eq:TT_formula2}
- \sum_{m,n=1}^\infty \sqrt{mn} \, (\gamma \alpha^{-1})_{m,n} z^{-m-1} w^{-n-1} =  \frac{f'(z)f'(w)}{(f(z)-f(w))^2} - \frac{1}{(z-w)^2}
\end{equation}
%
\end{prop}

\begin{proof}
Let $f=\sum_{n=1}^\infty f_n z^n$ with $f_1 \neq 0$. Observe that 
the expression  $f(z)^{n-1} f'(z)z^m$ in equation \eqref{eq:gamma} is regular at zero and its integral over $C$ vanishes. Hence, the operator $\gamma$ vanishes. Furthermore, for $n>m$ the function $f(z)^{n-1} f'(z)/z^m$ in equation \eqref{eq:alpha} is also regular at zero which implies $\alpha_{m,n}=0$. Therefore, $\alpha$ is an upper-triangular (infinite) matrix. A similar argument shows that $\delta$ is also upper-triangular.

Let $h$ be the inverse function of $f$. Equation \eqref{eq:alpha} implies
$$
\alpha^{-1}_{m,n} = \frac{1}{2\pi i} \, \sqrt{\frac{{n}}{{m}}} \, \int_{C'} \frac{h(w)^{n-1} h'(w)}{w^m} dw,
$$
where $C'$ is some (possibly different from $C$) circle around zero. 
Combining with equation \eqref{eq:beta}, we obtain
$$
(\alpha^{-1} \beta)_{m,n} = \sum_{k=1}^\infty \alpha^{-1}_{m,k} \beta_{k,n} =
- \frac{1}{(2\pi i)^2} \, \sqrt{\frac{n}{m}} \, \int_{C \times C'} \frac{h'(w) f'(z)}{w^m f(z)^{n+1}} \,
\sum_{k=1}^\infty \frac{h(w)^{k-1}}{z^k} dw dz.
$$
Summing up a geometric series and making a substitution $z = h(u)$ yields
$$
(\alpha^{-1} \beta)_{m,n} = - \frac{1}{(2\pi i)^2} \,\sqrt{\frac{n}{m}} \, \int_{C' \times C'} \frac{h'(w) }{w^m u^{n+1}} \, \frac{1}{h(u)-h(w)} \, du dw
$$
Finally, integration by parts over $u$ gives rise to
$$
(\alpha^{-1} \beta)_{m,n} = \frac{1}{(2\pi i)^2} \, \frac{1}{\sqrt{mn}} \, 
\int_{C' \times C'} \frac{1}{w^m u^{n}} \, \frac{h'(u)h'(w)}{(h(u)-h(w))^2} \, du dw.
$$
The function
$$
\frac{h'(u)h'(w)}{(h(u)-h(w))^2} - \frac{1}{(u-w)^2} = \frac{\partial^2}{\partial u \partial w} \, \log\left( \frac{f(u)-f(w)}{u-w}\right)
$$
is regular in $u,w$. Hence, it is given by the Taylor series \eqref{eq:TT_formula1}. 

Proof of equation \eqref{eq:TT_formula2} is similar.
\end{proof}

\section{Metaplectic representation and Berezin formalism}

In this Section, we recall the metaplectic representation of the symplectic group and Berezin formalism in finite and infinite dimensions.

\subsection{Heisenberg Lie algebra and normal symbols}
To a symplectic vector space $V \oplus V^*$ one can naturally associate a
 Heisenberg Lie algebra with generators 
$\hat{a}, \hat{a}^*$ for $a\in V, a^* \in V^*$ defined by canonical commutation relations
$$
[\hat{a}, \hat{b}] = [\hat{a}^*, \hat{b}^*]=0, \hskip 0.3cm
[ \hat{a}, \hat{b}^*] = \omega(b^*, a)= \langle b, a\rangle.
$$
Choose a Hermitian scalar product $(\cdot, \cdot)$ on $V$. Then, the symmetric algebra $SV^*$ also carries a Hermitian product, and it can be completed to a Fock space
$$
\mathcal{F} = \overline{SV^*}.
$$
The Fock space $\mathcal{F}$ carries a natural action (by unbounded operators) of the Heisenberg algebra, where $\hat{a} \cdot 1 = 0$ for all $a \in V$ and
$$
(\hat{a} + \hat{a}^*) \cdot f = a^* f + \partial_a f,
$$
where $\partial_a$ is a constant vector field acting on $SV^*$. 

Introduce an orthonormal basis $\{ a_i\}$ of $V$ and the dual basis $\{ a_i^* \}$ of $V^*$. In this basis, $\omega$ takes the canonical form
$$
\omega(a_i, a_j)=\omega(a^*_i, a^*_j)=0, \hskip 0.3cm
\omega(a_i, a^*_j) = \delta_{ij}.
$$
Then, operators $\hat{a}_i, \hat{a}^*_i$ on $\mathcal{F}$ are conjugate to each other under the Hermitian structure on $\mathcal{F}$.
To multi-indices $I=(i_1, \dots, i_m), J=(j_i, \dots, j_n)$ we associate
monomials
$$
{a}_I = {a}_{i_1} \dots {a}_{i_m}, \hskip 0.3cm
{a}_J={a}^*_{j_1} \dots {a}^*_{j_n}.
$$
To a power series in formal variables $a_i, a^*_i$
$$
N_q(a, a^*) = \sum_{I, J} q_{I,J} a_I a^*_J
$$
one associates an operator
$$
\hat{q} = \sum_{I, J} q_{I, J} \hat{a}^*_J \hat{a}_I.
$$
If the sum is finite, this operator is well defined, and $N_q(a, a^*)$ is called its normal symbol. Sometimes, $\hat{q}$ is well defined even for infinite series $N_q(a, a^*)$. The operator product $q \cdot r$ is represented by a formal Gaussian integral in terms of normal symbols:
\begin{equation} \label{eq:f*g}
N_q * N_r = N_{q \cdot r}(a, a^*) = \int  N_q(a+b, a^*) \, e^{-\langle b^*, b\rangle} \, 
N_r(a, a^* + b^*) \, dbdb^*.
\end{equation}
Note that this formal integral in defined modulo sign since in general it involves a square root of the determinant (see below for a more detailed discussion). 

\subsection{Berezin formalism}
In the finite dimensional context, the group of symplectic transformations
${\rm Sp}(V\oplus V^*)$ has a double cover
$$
1 \to \mathbb{Z}_2 \to {\rm Mp}(V \oplus V^*) \to {\rm Sp}(V \oplus V^*) \to 1
$$
called the metaplectic group. We will also need the associated central extension of ${\rm Sp}(V \oplus V^*)$ by $\mathbb{C}^*$:
$$
1 \to \mathbb{C}^* \to \widehat{\rm Sp}(V \oplus V^*)=
{\rm Mp}(V \oplus V^*) \times_{\mathbb{Z}_2} \mathbb{C}^*
\to {\rm Sp}(V \oplus V^*) \to 1.
$$
It comes with a natural representation  on the Fock space $\mathcal{F}$ which can also be viewed as a projective representation of ${\rm Sp}(V\oplus V^*)$. 

For a given symplectic transformation $A \in {\rm Sp}(V \otimes V^*)$
one says that an invertible operator $\hat{A}$ on $\mathcal{F}$ implements it if it represents a lift of $A$ in 
$\widehat{\rm Sp}(V\oplus V^*)$. In more detail, it means that
\begin{equation}
    \hat{A}
    \left(
    \begin{array}{l}
    \hat{a} \\
    \hat{a}^*
    \end{array}
    \right) 
    \hat{A}^{-1}
    =
    \left(
    \begin{array}{ll}
    \alpha & \beta \\
    \gamma & \delta
    \end{array}
    \right)
    \left(
    \begin{array}{l}
    \hat{a} \\
    \hat{a}^*
    \end{array}
    \right)
    \end{equation}
    for all $a \in V, a^* \in V^*$.

We will call a symplectic transformation $A \in {\rm Sp}(V \oplus V^*)$ admissible if its components $\alpha: V \to V$ and $\delta: V^* \to V^*$ are invertible.
The following theorem summarizes a result of Berezin (see \cite{Berezin}):
\begin{thm}  \label{thm:berezin}
Let $A \in {\rm Sp}(V \oplus V^*)$ be an admissible symplectic transformation. Then, $A$ is implemented by a unique operators $\hat{A}$ with a normal symbol $N_A$ whose constant term is equal to 1. This normal symbol is given by formula
\begin{equation}   \label{eq:normal_symbols}
N_A(a, a^*)  =     
\exp\left( \langle a^*, (\alpha^{-1} -1) a\rangle - \frac{1}{2} (a^*, (\alpha^{-1} \beta) a^*) + \frac{1}{2} (a, (\gamma \alpha^{-1}) a) \right)    
\end{equation}
\end{thm}

We will call a pair of admissible symplectic transformations $A_1, A_2$ composable if $A_1A_2$ is also an admissible transformation. We use a similar terminology for triples.
The following proposition gives a product rule in terms of normal symbols:

\begin{prop}  \label{prop:berezin_product}
Let $A_1, A_2 \in {\rm Sp}(V \oplus V^*)$ be a composable pair. 
Then,
\begin{equation}   \label{eq:berezin_product}
N_{A_1} * N_{A_2} = \frac{1}{{\rm det}^{1/2}(1 + (\alpha_2^{-1} \beta_2) (\gamma_1 \alpha_1^{-1}))} \, N_{A_2A_1}.
\end{equation}
\end{prop}

\begin{proof}
The proof is by a direct calculation of the Gaussian integral \eqref{eq:f*g}.
\end{proof}

Note that the product rule for normal symbols \eqref{eq:berezin_product} is not quite well defined because of the square root of the determinant. In fact, the subset of admissible elements in the metaplectic group admits the following description:
$$
{\rm Mp}_{\rm adm}(V \oplus V^*) = \{ (A, z) \in {\rm Sp}_{\rm adm}(V \oplus V^*) \times \mathbb{C}^*; {\rm det}(\alpha) = z^2\}.
$$
Observe that
$$
{\rm det}(1 + (\alpha_2^{-1} \beta_2)( \gamma_1 \alpha_1^{-1}))
= \frac{{\rm det}(\alpha)}{{\rm det(\alpha_1)} {\rm det}(\alpha_2)}.
$$
Hence, we can re-write the product rule of normal symbols in terms of the metaplectic group as follows:
$$
N_{(A_1, z_1)} * N_{(A_2, z_2)} = \frac{z_1z_2}{z} N_{(A, z)},
$$
where $A=A_2A_1$.
One can summarize the properties of the product rule as follows:
\begin{prop} 
The expression
\begin{equation}   \label{eq:berezin_cocycle}
C_N(A_2, z_2; A_1, z_1) = \frac{1}{{\rm det}^{1/2}(1 + (\alpha_2^{-1} \beta_2) (\gamma_1 \alpha_1^{-1}))} = \frac{z_1z_2}{z}
\end{equation}
is a multiplicative 2-cocylce. That is, for all composable triples $A_1, A_2, A_3$ we have
$$
C_N(A_2, A_1) C_N(A_3, A_2A_1) = C_N(A_3A_2, A_1)C_N(A_3, A_2)
$$
\end{prop}

\begin{proof}
The statement follows from the fact that $C_N=z_1z_2/z$ is a trivial 2-cocylce. This fact reflects  associativity of the operator product $*$.
\end{proof}

For operators with normal symbols $G_{(A,z)}=z^{-1} N_A$, we obtain a group anti-homomorphism:
\begin{equation}
    G_{(A_2, z_2)} * G_{(A_1, z_1)} = G_{(A_1A_2, z_1z_2)}.
\end{equation}

The group ${\rm Sp}(V \oplus V^*)$ contains a subgroup ${\rm USp}(V \oplus V^*)$ which has the following property: $\tilde{a}_i$ is the conjugate of $\tilde{a}^*_i$ for all $i$. This condition imposes an extra requirement on the components $\alpha, \beta, \gamma, \delta$ of $A$:
$$
\gamma = \overline{\beta}, \hskip 0.3cm 
\delta=\overline{\alpha}.
$$
Here $\overline{\alpha}, \overline{\beta}$ are complex conjugate of $\alpha$ and $\beta$, respectively. In particular, we obtain
$$
\alpha \alpha^* = \alpha \delta^t = 1 + \beta \gamma^t = 1 + \beta \beta^*.
$$
This implies that $\alpha$ is invertible, and that so is $\delta = \overline{\alpha}$. Furthermore, we have the following useful identity:
and
\begin{equation}   \label{eq:alpha_alpha_-1}
\alpha^{-1}(\alpha^*)^{-1} = \alpha^{-1}(\alpha \alpha^* - \beta \beta^*)(\alpha^*)^{-1} = 1 - (\alpha^{-1} \beta)(\alpha^{-1}\beta)^*.
\end{equation}
This implies that all transformations $A \in {\rm USp}(V \oplus V^*)$ are admissible, all pairs $A_1, A_2$ are composable, and Theorem \ref{thm:berezin} and Proposition \ref{prop:berezin_product} apply without further assumptions. Also, the map $(A, z) \mapsto G_{(A,z)}$ defines a group anti-homomorphism from the corresponding subgroup of the metaplectic group ${\rm MUSp}(V \oplus V^*)$ to unitary operators on the Fock space $\mathcal{F}$.

\subsection{The infinite dimensional case}
Most of the facts reviewed in the previous Section generalize to the infinite dimensional setup. Let $V$ be a Hilbert space. This allows to identify $V^* \cong V$. In what follows, we list special features which distinguish the infinite dimensional situation from the finite dimensional one.

Instead of the symplectic group ${\rm Sp}(V \oplus V^*)$, one considers the restricted symplectic group
$$
{\rm Sp}^{\rm res}(V \oplus V^*) = \{ A \in {\rm Sp}(V \oplus V^*); \alpha, \delta \, {\rm are} \, {\rm Fredholm},
\beta, \gamma \, {\rm are}\, {\rm Hilbert-Schmidt}\}.
$$
For admissible elements of this subgroup, Theorem \ref{thm:berezin} and Proposition \ref{prop:berezin_product} hold true verbatim. In particular, the determinant
$$
{\rm det}(1 + (\alpha_2^{-1} \beta_2) (\gamma_1 \alpha_1^{-1}))
$$ 
is well defined since the operator
$$
1 + (\alpha_2^{-1} \beta_2) (\gamma_1 \alpha_1^{-1}) = \alpha_2^{-1} \alpha \alpha_1^{-1}
$$
is invertible, the operators $(\alpha_2^{-1} \beta_2)$ and $(\gamma_1 \alpha_1^{-1})$ are Hilbert-Schmidt, and hence the operator $(\alpha_2^{-1} \beta_2) (\gamma_1 \alpha_1^{-1})$ is of trace class. The product formula \eqref{eq:berezin_product} still makes sense on the metaplectic double cover.

However, the determinant ${\rm det}(\alpha)$ is not well defined, in general. Therefore, the admissible part of the metaplectic group does not allow for a simple description using the equation $z^2={\rm det}(\alpha)$, and the cocycle $C_N(A_1, A_2)$ is {\em a priori} nontrivial. It is instructive to write the corresponding Lie algebra cocycle
$a_N(x_1, x_2)$ on a pair of elements of the symplectic Lie algebra:
$$
x_i =
\left(
\begin{array}{ll}
a_i & b_i \\
c_i & d_i
\end{array}
\right), \, i=1,2,
$$
where $a_i$ and $d_i$ are bounded operators, and $b_i$ and $c_i$ are Hilbert-Schmidt operators. 
An easy calculation shows that
$$
a_N(x_1, x_2) = {\rm Tr}(b_1c_2 -c_1b_2).
$$
The right hand side is well defined because both terms $b_1c_2$ and $c_1b_2$ are of trace class.

The restricted group ${\rm USp}^{\rm res}(V \oplus V^*)$ is defined as before:
$$
{\rm USp}^{\rm res}(V \oplus V^*) = \{ A \in {\rm Sp}^{\rm res}(V \oplus V^*); \delta=\overline{\alpha}, \gamma=\overline{\beta}\} .
$$
Again, all elements $A \in {\rm USp}^{\rm res}(V \oplus V^*) $ are admissible and all pairs $A_1, A_2$ are composable. 

Another important result of \cite{Berezin} is as follows:
\begin{thm}  \label{prop:berezin_unitary}
Let $A \in {\rm USp}^{\rm res}(V \oplus V^*)$. Then, the normal symbol
$$
U_A = \pm \frac{1}{{\rm det}(\alpha \alpha^*)^{1/4}} \, N_A
= \pm {\rm det}^{1/4}(1 - (\alpha^{-1}\beta)(\alpha^{-1}\beta)^*) \, N_A
$$
defines a unitary operator on $\mathcal{F}$.
\end{thm}
Here we have used equation \eqref{eq:alpha_alpha_-1}. Note that the
resulting Fredholm determinant is well defined since the operator 
$(\alpha^{-1}\beta)(\alpha^{-1}\beta)^*$ is of trace class.
Operators $U_A$ satisfy the product rule
$$
U_{A_1} * U_{A_2} = C_U(A_1, A_2) U_{A_2 A_1},
$$
where the cocycle $C_U(A_1, A_2)$ is given by formula
$$
C_U(A_1, A_2) = \left(\frac{{\rm det}(\alpha \alpha^*)}{
{\rm det}(\alpha_1 \alpha_1^*){\rm det}(\alpha_2 \alpha_2^*)}\right)^{\frac{1}{4}} \, \cdot \, \frac{1}{{\rm det}^{1/2}(1 + (\alpha_2^{-1} \beta_2)( \gamma_1 \alpha_1^{-1}))}.
$$
It is defined on the metaplectic double cover, and it
takes values in $S^1 \cong \{ z\in \mathbb{C}; |z|=1\}$ instead of $\mathbb{C}^*$. This cocycle is non trivial, in general. The corresponding Lie algebra cocycle is the same as for $C_N(A_1, A_2)$ (up to second order, the normalization factor is symmetric in $A_1, A_2$):
$$
a(x_1, x_2) = {\rm Tr}(b_1 \overline{b}_2 - \overline{b}_1 b_2).
$$

\section{Quantization of conformal welding}

In this Section, we apply Berezin quantization to triangular decomposition of symplectic transformations induced by holomorphic maps, and in particular to conformal welding.

\subsection{Triangular decomposition and conformal welding}
In this Section, we discuss an analogue of triangular decomposition for holomorphic maps.

Let $f: \mathcal{A}_{r, R} \to \mathbb{C}$ be a univalent holomorphic map defined on the annulus $\mathcal{A}_{r, R}$ such that its range is contained in another annulus: $f(\mathcal{A}_{r, R}) \subset \mathcal{A}_{r', R'}$. We say that $f$ admits a triangular decomposition if there exist univalent holomorphic maps $f_+: \mathbb{D}_{R'} \to \mathbb{C}, f_-: \mathbb{D}^*_{r'}$ such that $f_+(0)=0, f_-(\infty)=\infty$ and
\begin{equation}  \label{eq:triangular}
f_- = f_+ \circ f
\end{equation}
on $\mathcal{A}_{r, R}$. Here
$$
\mathbb{D}_R=\{ z \in \mathbb{C}; |z|<R\}, \hskip 0.3cm
\mathbb{D}^*_r = \{ z \in \mathbb{C}; |z| >r\} \cup \{ \infty\}
$$
are the discs centered at $0$ and $\infty$, respectively.
Holomorphic functions $f_+$ and $f_-$ admit Taylor expansions at $0$ and $\infty$: 
\begin{equation}   \label{eq:f_pm}
f_+(z) = \sum_{k=1}^\infty (f_+)_k z^k, \hskip 0.3cm
f_-(z) = \sum_{k=-\infty}^1 (f_-)_k z^k,
\end{equation}
where $(f_+)_1 \neq 0, (f_-)_1 \neq 0$. 

A special case of triangular decomposition is given by conformal welding of diffeomorphisms of the circle. Let $f$ be a holomorphic function which corresponds to $\chi \in {\rm Diff}^+_{\rm hol}(S^1)$. That is,
$$
f(e^{ix})= e^{i \chi(x)}.
$$
In this case, one can choose the domain of $f$ to be an annulus $\mathcal{A}_{r, R}$ with $r < 1 < R$. The following theorem follows from results of \cite{Kirillov_welding} on conformal welding for elements of
${\rm Diff}^+(S^1)$. We will only be interested in the subgroup ${\rm Diff}^+_{\rm hol}(S^1) \subset{\rm Diff}^+(S^1)$.
\begin{thm}
Let $f: \mathcal{A}_{r, R} \to \mathbb{C}$ be a univalent holomorphic map which corresponds to a diffeomorphism of the circle $\chi$. Then, it admits a unique triangular decomposition $f =f_+^{-1} \circ f_-$ with $(f_+)'(0)=1$.
\end{thm}

\begin{rem}
One says that the univalent holomorphic functions $f_\pm$ provide a conformal welding of the diffeomorphism $\chi$.
Note that the normalization $f'_+(0)=(f_+)_1=1$ (here $f_+(z)=\sum_{n=1}^\infty (f_+)_n z^n$)
can be replaced by the normalization $f'_-(\infty)=(f_-)_1 =1$ (here $f_-(z) =
\sum_{n=-\infty}^1 (f_-)_n z^n$. This is achieved by dividing both $f_+$ and $f_-$ by $f'_-(\infty)$.
\end{rem}

Equation \eqref{eq:triangular} can also be re-written in the form
$$
f = f_+^{-1} \circ f_-
$$
which resembles of the Gauss decomposition of matrices. It turns out that it induces a Gauss decomposition on the corresponding elements of the infinite dimensional symplectic group. Indeed, by Proposition \ref{prop:f_h},  symplectic transformations $A_{f_+}$ and $A_{f_-}$ are upper- and lower-triangular, respectively. And by Proposition \ref{prop:A_fg}, we have
a Gauss type decomposition (recall that the map $f \mapsto A_f$ is a group anti-homomorphism):
$$
A_f = A_{f_-} A_{f_+^{-1}}.
$$
It is convenient to introduce the notation
$$
A_f =
\left(
\begin{array}{ll}
\alpha_f & \beta_f \\
\gamma_f & \delta_f
\end{array}
\right)
$$
for components of the symplectic transformation $A_f$. Then, we have
$$
A_{f} = 
\left(
\begin{array}{ll}
\alpha_{f_-} &  0 \\
\gamma_{f_-} & \delta_{f_-}
\end{array}
\right)
\left(
\begin{array}{ll}
\alpha_{f_+^{-1}} & \beta_{f_+^{-1}} \\
0 & \delta_{f_+^{-1}}
\end{array}
\right) =
\left(
\begin{array}{cc}
\alpha_{f_-} \alpha_{f_+^{-1}}& \alpha_{f_-} \beta_{f_+^{-1}} \\
\gamma_{f_-} \alpha_{f_+^{-1}} & \gamma_{f_-} \beta_{f_+^{-1}} + \delta_{f_-} \delta_{f_+^{-1}}
\end{array}
\right).
$$
Assuming that $\alpha_{f_-}$ and $\alpha_{f_+}^{-1}$ are invertible, this implies
\begin{equation}  \label{eq:Gauss}
    \alpha_f^{-1} \beta_f = \alpha_{f_+^{-1}}^{-1}\beta_{f_+^{-1}}, \hskip 0.3cm
    \gamma_f \alpha_f^{-1} = \gamma_{f_-} \alpha_{f_-}^{-1}.
\end{equation}

\subsection{Normal symbols of holomorphic maps}
In this Section, we apply Berezin theory of normal symbols to holomorphic maps and the corresponding symplectic transformations.

In order to do that, we equip the space of holomorphic functions (modulo constants) $H$ with a structure of a Hilbert space (following \cite{Sullivan}) by declaring $|| z^n || = |n|$ for all $n \neq 0$. Then, symplectic transformations induced by holomorphic maps admit a metaplectic projective representation on the corresponding Fock space.

In more detail, let $f$ be a holomorphic map, and assume that the corresponding symplectic transformation $A_f$ is admissible. Then, it is convenient to denote by $N_f$ (instead of $N_{A_f}$) its normal symbol.
\begin{prop}  \label{prop:N*N}
Let $f$ and $g$ be a pair of composable holomorphic maps, and assume that the corresponding symplectic transformations $A_f$ and $A_g$ are admissible. Then,
$$
N_f * N_g = \frac{1}{{\rm det}^{1/2}(1 + (\alpha_g^{-1}\beta_g)(\gamma_f \alpha_f^{-1}))} \, N_{f\circ g}
$$
Furthermore, if $f=\sum_{n=1}^\infty f_n z^n$ with $f_1 \neq 0$, or if 
$g=\sum_{-\infty}^1 g_n z^n$ with $g_1 \neq 0$, then
$$
N_f * N_g = N_{f\circ g}.
$$
\end{prop}

\begin{proof}
The first statement follows from Proposition \ref{prop:berezin_product}. 
For the second statement, note that if $f=\sum_{n=1}^\infty f_n z^n$, then
by Proposition \ref{prop:f_h} $\gamma_f=0$. Similarly, if $g=\sum_{-\infty}^1 g_n z^n$, then $\beta_g=0$. This completes the proof.
\end{proof}

We are now ready to state one of our main results:
\begin{thm}   \label{thm:C_N}
Let $f$ and $g$ be a pair of composable holomorphic maps, and assume that they admit triangular decompositions. Then,
\begin{equation} \label{eq:product_N}
N_{f} * N_{g} = C_N(f, g) N_{f \circ g},
\end{equation}
where the 2-cocycle $C_N(f,g)$ is of the form
\begin{equation}  \label{eq:C_N}
    C_N(f,g)=\frac{1}{ {\rm det}^{1/2}(1 + (\alpha_{g_+^{-1}}^{-1}\beta_{g_+^{-1}})(\gamma_{f_-}\alpha_{f_-^{-1}}))}.
\end{equation}
The corresponding map $\beta_N$ is given by formula
\begin{equation}  \label{eq:beta_N}
    \beta_N(u, g) = - \frac{1}{(2\pi i)^2} 
    \int_{C \times C} \frac{u_-(z)-u_-(w)}{z-w} \, \frac{\partial^2}{\partial z \partial w} \, \log\left( \frac{g_+(z)-g_+(w)}{z-w} \right) \, dzdw.
\end{equation}

\end{thm}

\begin{proof}
For the first statement, we use Proposition \ref{prop:N*N}, and we observe that by formula \eqref{eq:Gauss}
$$
\alpha_g^{-1}\beta_g = \alpha_{g_+^{-1}} \beta_{g_+^{-1}}, \hskip 0.3cm
\gamma_f \alpha_f^{-1} = \gamma_{f_-} \alpha_{f_-}^{-1}.
$$
In order to prove the formula for $\beta_N$, put $f(t)=\exp(tu)$, where 
$u=u(z)\frac{\partial}{\partial z}$ is a vector field with 
$$
u(z) = u_+(z) + u_-(z) = \sum_{n=1}^\infty u_n z^n + \sum_{n=-\infty}^0 u_n z^n.
$$
Note that $f(t)=f_+(t)^{-1} f_-(t)$ and $\log(f_-(t)) = tu_- + O(t^2)$. Hence,
$$
\beta_N(u, g) = \frac{d}{dt} \, C(\exp(tu), g)|_{t=0} =
{\rm Tr} (\alpha_{g+^{-1}}^{-1} \beta_{g_+^{-1}}) \left(\frac{d}{dt} \, (\gamma_{\exp(tu_-)} \alpha_{\exp(tu_-)}^{-1})\right)_{t=0}.
$$
Recall that 
$$
\sum_{m,n} \sqrt{mn} (\gamma_{\exp(tu_-)} \alpha_{\exp(t u_-)}^{-1})_{m,n} z^{-m-1}w^{-n-1} = 
- \frac{\partial^2}{\partial z \partial w} \, \log\left(\frac{\exp(tu_-)(z)-\exp(tu_-)(w)}{z-w}\right),
$$
where $\exp(tu_-)(z)$ is the image of $z$ under the holomorphic map $\exp(tu_-)$. The derivative in $t$ at $t=0$ yields (after integrating over $z$ and $w$)
$$
\sum_{m,n} \frac{1}{\sqrt{mn}} \, \frac{d}{dt}(\gamma_{f_-} \alpha_{f_-}^{-1})_{m,n}|_{t=0} z^{-m}w^{-n} = 
- \frac{u_-(z)-u_-(w)}{z-w}.
$$
Also recall that
$$
\sum_{m,n} \sqrt{mn} (\alpha_{g_+^{-1}} \beta_{g_+})_{m,n} z^{m-1}w^{n-1}=
\frac{\partial^2}{\partial z \partial w} \, \log\left( \frac{g_+(z)-g_+(w)}{z-w}\right).
$$
Next, we convert the trace in $m,n$ into a double contour integral.
Since the factors $\sqrt{mn}$ and $1/\sqrt{mn}$ cancel out, we obtain the desired result.
\end{proof}

Surprisingly, the cocycle
$$
C_N(f,g)=C_N(f_+^{-1} f_-, g_+^{-1} g_-)
$$
has the following polarization property: it is independent of the components $f_+$ and $g_-$ in triangular decompositions of $f$ and $g$.

%
%

The following result of \cite{TT} (see Corollary 2.9 in Chapter 2) establishes important properties of symplectic transformations associated to conformal welding:
\begin{thm}  \label{thm:TT1}
Let $\chi \in {\rm Diff}^+_{\rm hol}(S^1)$, $f$ the corresponding holomorphic function, and
$f_\pm$ the components of conformal welding of $\chi$.
Then, the maps $A_{f_+^{-1}}, A_{f_-}$ and $A_f = A_{f_-} A_{f_+^{-1}}$ belong to the restricted symplectic group ${\rm Sp}^{\rm res}(H_+ \oplus H_-)$. In particular, symmetric operators 
$(\alpha_{f_+^{-1}}^{-1} \beta_{f_+^{-1}})$ and $(\gamma_{f_-} \alpha_{f_-}^{-1})$ are Hilbert-Schmidt.
\end{thm}

Theorem \ref{thm:TT1} implies that conditions of Proposition \ref{prop:N*N} are verified for conformal welding, and we have
$$
N_f = N_{f_+^{-1}} * N_{f_-}.
$$

Furthermore, let  $\chi, \phi \in {\rm Diff}^+_{\rm hol}(S^{1})$. Recall that the correpsonding 
conformal maps $f$ and $g$ are always composable. By Theorem \ref{thm:TT1}, $A_f$ and $A_g$ are admissible. Hence, Theorem \ref{thm:C_N} applies and we obtain the formula
\eqref{eq:product_N} for the product of normal symbols $N_f * N_g$.

\subsection{The Takhtajan-Teo energy functional}

In this Section, we  recall the definition and main properties of the Takhtajan-Teo (TT) energy functional which will be important in the discussion of unitary symbols for conformal welding.

Let $\mathcal{D} \subset \mathbb{C}$ be a simply connected domain on the complex plane, $C \subset \mathcal{D}$ be a simple closed curve and $
\mathcal{C} \subset \mathcal{D}$ be a compact domain bounded by $C$.
We consider $u: \mathcal{D} \to \mathbb{C}$ a univalent holomorphic function. 
To a pair $(\mathcal{C}, u)$ we associate a functional
\begin{equation}
    E(\mathcal{C}, u) = \int_\mathcal{C} \, \left| \frac{u''(z)}{u'(z)} \right|^2 d^2z.
\end{equation}

Two pairs $(u, \mathcal{C}_u)$ and $(v, \mathcal{C}_v)$ are called compatible if $v(C_v)=C_u$ and  $v(\mathcal{C}_v) = \mathcal{C}_u$. 

\begin{prop}   \label{prop:TT_tranform}
For compatible pairs $(u, \mathcal{C}_u)$ and $(v, \mathcal{C}_v)$, we have
\begin{equation}
    E(u \circ v) = E(u) + E(v) -
    2\,  {\rm Re} \, \int_{\mathcal{C}_u} \, \frac{u''(w)}{u'(w)} \cdot \frac{\overline{(v^{-1})''(w)}}{\overline{(v^{-1})'(w)}} \, d^2w.
\end{equation}
\end{prop}

\begin{proof}
By direct substitution, we obtain
$$
\frac{(u \circ v)''(z)}{(u \circ v)'(z)} = 
\frac{u''(v(z))}{u'(v(z))} \, u'(z) + \frac{v''(z)}{v'(z)}.
$$
This implies
$$
E(u \circ v) = E(u) + E(v) + 2 \, {\rm Re} \, 
\int_{\mathcal{C}_v} \, \frac{u''(v(z))}{u'(v(z))} \, u'(z) \cdot 
\frac{\overline{v''(z)}}{\overline{v'(z)}} \, d^2z.
$$
We now make a change of variables $z=v^{-1}(w)$ in the last term, and use that
$(v^{-1})'(w)=u'(z)^{-1}$ and $(v^{-1})''(w)=-u''(z)/u'(z)^3$ to complete the proof.
\end{proof}

As an example, consider the case where $\mathcal{C}_u = \mathcal{C}_v=\mathbb{D}$ is the unit disk. In this case, we denote
$$
E_+(v) = E(\mathbb{D}, v).
$$
Let $v$ be a M\"obius transformation:
$$
v(z)= \frac{az+b}{\bar{b} z + \bar{a}},
$$
with $|a|^2 - |b|^2=1$. Then,
$$
E_+(v) = 4 |b|^2 \int_\mathbb{D} \, \frac{d^2z}{|\bar{b} z + \bar{a}|^2} =
- 4 \pi \, \log\left(1 - \left| \frac{b}{a} \right|^2 \right) = 4\pi \log(|a|^2).
$$
The inverse M\"obius transformation $v^{-1}(w)$ is of the form
$$
v^{-1}(w)=\frac{\bar{a} w - b}{a - \bar{b}w},
$$
and we have
$$
\begin{array}{lll}
\int_{\mathbb{D}} \, \frac{u''(w)}{u'(w)} \cdot \frac{\overline{(v^{-1})''(w)}}{\overline{(v^{-1})'(w)}} \, d^2w & = &
2\pi\left( \log(u'(b/\bar{a})) - \log(u'(0)) \right) \\
& = &
2\pi(\log((u\circ v)'(0)) - \log(u'(0)) + \log(\bar{a}^2)).
\end{array}
$$
Here we have used that $v(0)=b/\bar{a}$ and that
$(u \circ v)'(0) = u'(v(0))v'(0) = u'(v(0))/\bar{a}$.
Putting things together, we obtain the following interesting equality for the functional $E_+$:
$$
E_+(u \circ v) + 4\pi \log|(u \circ v)'(0)| = E_+(u) + 4\pi \log|u'(0)|.
$$
Hence, on the unit disk the expression
$$
S_+(u) = E_+(u) +4\pi \log(|u'(0)|)= \int_\mathbb{D} \left| \frac{u''(z)}{u'(z)} \right|^2 \, d^2z 
+ 4\pi \log(|u'(0)|)
$$
is invariant under M\"obius transformations. Similarly, for a function $u$ defined on the unit disk $\mathbb{D}^*$ centered at infinity, we define
$$
S_-(u) = E_-(u) - 4\pi \log(|u'(\infty)|)= \int_{\mathbb{D}^*} \left| \frac{u''(z)}{u'(z)} \right|^2 \, d^2z 
- 4\pi \log(|u'(\infty)|).
$$

Let $\chi \in {\rm Diff}^+_{\rm hol}$, $f$ the corresponding holomorphic map and $f=f_+^{-1} \circ f_-$ the  conformal welding of $\chi$. The TT energy functional is defined as
\begin{equation}
    S(\chi) = S_+(f_+) + S_-(f_-).
\end{equation}
By the invariance properties of $S_\pm$,
$S(\chi)$ is left- and right-invariant under M\"obius transformations:
$$
S(m_1 \circ \chi \circ m_2) = S(\chi).
$$
In particular, it descends to the quotient space
$$
{\rm Diff}^+_{\rm hol}(S^1)/{\rm PSL}(2, \mathbb{R}) \subset {\rm Diff}^+(S^1)/{\rm PSL}(2, \mathbb{R}).
$$

We recall the following highly nontrivial property of the TT energy functional (see Theorem 3.8 in Chapter 2 of \cite{TT}):
\begin{thm} \label{thm:S_symmetry}
For all $\chi \in{\rm Diff}^+(S^1)$, we have
$$ S(\chi^{-1}) = S(\chi).$$
\end{thm}

We observe the following interesting property of the functional $S(\chi)$:
\begin{prop}  \label{prop:TT=cocycle+}
For $\chi \in {\rm Diff}^+_{\rm hol}$, we have
$$
S(\chi) =  {\rm Im} \, \tilde{C}(f_-^{-1}, f_+) - \int_{C} (\chi'(x) + 1) \log(\chi'(x)) dx,
$$
where $\tilde{C}$ is the groupoid 2-cocycle defined by equation \eqref{eq:tildeC}.
\end{prop}

\begin{proof}
First, observe that
$$
E_+(f_+) = \frac{i}{2} \, \int_{C} \,  \log(f'_+) d \log(\overline{f'_+}), \hskip 0.3cm
E_-(f_-) = - \frac{i}{2} \, \int_{C} \, \log(f'_-) d \log(\overline{f'_-}).
$$
Also, note that 
$$
{\rm Im}\, \tilde{C}(f_-^{-1}, f_+) = {\rm Im} \,  \tilde{C}(f_+, f) = - \frac{i}{2} \int_C (\log(f'_+(f)) d \log(f') - \log(\overline{f'_+(f)}) d \log(\overline{f'})).
$$
We consider
$$
\begin{array}{lll}
A_1(\chi)  & = & 
E_+(f_+) + E_-(f_-) - {\rm Im} \,  \tilde{C}(f_-^{-1}, f_+) \\
& = &  \frac{i}{2} \, \int_{C} \,  \log(f'_+) d \log(\overline{f'_+}) \\
& - & \frac{i}{2} \, \int_{C} \, (\log(f'_+(f) + \log(f')) d (\log(\overline{f'_+(f)}) + \log(\overline{f'})) \\
& + & \frac{i}{2} \int_C (\log(f'_+(f)) d \log(f') - \log(\overline{f'_+(f)}) d \log(\overline{f'}))
\end{array}
$$
Using that for $z=e^{ix}$ 
$$
\log(f'(z))=\log(\chi'(x)) + i (\chi(x)-x),
$$
we obtain
$$
A_1(\chi) = - \int_C (\log(\chi'(x)) + \log(|f'_+(f(z))|^2)(\chi'(x)-1) dx.
$$
Next, we consider 
$$
\begin{array}{lll}
A_2(\chi) & = &
4\pi(\log(|f'_+(0)|) - \log(|f'_-(\infty)|) \\
& = & \int_C (\log(|f'_+(z)|^2) - \log(|f'_-(z)|^2)) dx \\
& = & \int_C (\log(|f'_+(z)|^2) - \log(|f'_+(f(z))|)^2 - \log(|f'(z)|^2)) dx \\
& = & \int_C (\log(|f'_+(f(z))|)^2(\chi'(x) -1) - 2 \log(\chi'(x))) dx.
\end{array}
$$
Adding up the expressions $A_1(\chi)$ and $A_2(\chi)$, we conclude
$$
S_+(f_+) + S_-(f_-) - {\rm Im} \, \tilde{C}(f_-^{-1}, f_+) = - \int_C (\chi'(x) + 1)\log(\chi'(x)) dx,
$$
as required.
\end{proof}

Note that the statement of Proposition \ref{prop:TT=cocycle+} can be re-written as
\begin{equation}  \label{eq:tildeC=}
{\rm Im} \, \tilde{C}(f_-^{-1}, f_+)=S(\chi)  + \int_{C}  (\log(\chi'(x)) - \log((\chi^{-1})'(x))) dx.
\end{equation}
By Theorem \ref{thm:S_symmetry}, the first term on the right hand side is invariant under the involution $\chi \mapsto \chi^{-1}$ 
while the second term is anti-invariant.

\subsection{Unitary symbols and quantization of welding}
In this Section, we study properties of unitary symbols $U_f$ corresponding to diffeomorphisms of the circle using conformal welding.

%
%

Let $\chi \in {\rm Diff}^+_{\rm hol}(S^1)$ and $f$ the corrsponding holomorphic map. Recall that the symplectic transformation $A_f \in {\rm USp}^{\rm res}(H_+ \oplus H_-)$ belongs to the unitary symplectic group. This implies that $\alpha_f$ is invertible, $\gamma_f = \bar{\beta}_f$, and %
$$
\alpha_f^{-1} (\alpha_f^*)^{-1} = \alpha_f^{-1}(\alpha_f\alpha_f^* - \beta_f \beta_f^*) (\alpha_f^*)^{-1} = 1 - (\alpha_f^{-1} \beta_f)(\alpha_f^{-1} \beta_f)^*.
$$
Since the symmetric operator $\alpha_f^{-1} \beta_f = \alpha_{f_+^{-1}}^{-1} \beta_{f_+^{-1}}$ is Hilbert-Schmidt, the operator $(\alpha_{f}^{-1} \beta_f)(\alpha_f^{-1} \beta_f)^*$ is trace class and $\alpha_f^{-1} (\alpha_f^*)^{-1}$ possesses a Fredholm determinant. The following result is an adaptation of Theorem 3.8 in Chapter 2 of \cite{TT}:
\begin{thm}  \label{thm:TT2}
The operator $\alpha_f$ is bounded and invertible. The Fredholm determinant of the self-adjoint operator $\alpha_f^{-1} (\alpha_f^*)^{-1}$ is given by
\begin{equation}
{\rm det}(\alpha_f^{-1} (\alpha_f^*)^{-1}) = e^{-S(\chi)/12\pi},
\end{equation}
where $S(f)$ is the TT energy functional.
\end{thm}

By Proposition \ref{prop:berezin_unitary}, the unitary operator representing the symplectic transformation  $A_f$ is given by
\begin{equation}  \label{eq:U_f}
U_f = \frac{1}{{\rm det}^{1/4}(\alpha_f \alpha_f^*)} \, N_f = e^{S(\chi)/48 \pi} \, N_f = e^{S(\chi)/48\pi } \, N_{f_+^{-1}} * N_{f_-}.
\end{equation}
Using this construction, we obtain the following interesting relation between the cocycle $C_N$ and the TT energy functional:
\begin{thm}  \label{thm:C_N_welding}
Let $\chi \in {\rm Diff}^+_{\rm hol}(S^1)$, $f$ the corresponding holomorphic function, and $f_\pm$ the components of the conformal welding of $\chi$.
Then,
$$
\log(|C_N(f_-^{-1}, f_+)|) = \frac{S(\chi)}{24\pi}.
$$
\end{thm}

\begin{proof}
On the one hand,  equation \eqref{eq:U_f} implies:
$$
U_f^{-1} = e^{- S(\chi)/48\pi } \, N_{f_-}^{-1} * N_{f_+^{-1}}^{-1} = e^{- S(\chi)/48\pi } \, N_{f_-^{-1} } * N_{f_+} = e^{- S(\chi)/48 \pi} \, C_N(f_-^{-1}, f_+)  N_{f^{-1}}.
$$
Here we have used the facts that $N_{f_-^{-1}}$ is the operator inverse of $N_{f_-}$ and $N_{f_+^{-1}}$ is the operator inverse of $N_{f_+}$.
On the other hand, we have
$$
U_{f^{-1}} = e^{S(\chi^{-1})/48 \pi} \, N_{f^{-1}}.
$$
Since both $U_f$ and $U_{f^{-1}}$ are unitary operators, their product $U_fU_{f^{-1}} = z \cdot  {\rm Id}$ which implements $f \circ f^{-1} = e$ is a multiple of the identity with $|z|=1$.
Hence,
$$
U_{f^{-1}} = z \, U_f^{-1} = z \, e^{- S(\chi)/48 \pi} \, C_N(f_-^{-1}, f_+)  N_{f^{-1}}.
$$
By comparing the two expressions for $U_{f^{-1}}$, we obtain the following equality:
$$
\log(|C_N(f_-^{-1}, f_+)|) = \frac{S(\chi) + S(\chi^{-1})}{48\pi} = \frac{S(\chi)}{24\pi},
$$
as required.
\end{proof}

\begin{rem}
The statement of Theorem \ref{thm:C_N_welding} should be compared to Proposition \ref{prop:TT=cocycle+} and equation
\eqref{eq:tildeC=}.
In particular, it would be interesting to find a more explicit relation between the cocycles
$\log(|C_N|)$ and $\tilde{C}$.
\end{rem}

The second main result of this article is the following theorem:
\begin{thm}  \label{thm:main2}
Let $\chi, \phi \in{\rm Diff}^+_{\rm hol}(S^1)$, and  $f$ and $g$ be the corresponding holomorphic maps. Then,
$$
U_f * U_g = C_U(f,g) \, U_{f \circ g},
$$
where 
\begin{equation}  \label{thm:C_U}
    C_U(f,g) = e^{(S(\chi) + S(\phi) - S(\chi \circ \phi))/48 \pi} \, C_N(f,g).
\end{equation}
Furthermore,
\begin{equation}  \label{eq:C_N=TT}
\log(|C_N(f,g)|) = \frac{S(\chi \circ \phi)-S(\chi)-S(\phi)}{48 \pi}.
\end{equation}
\end{thm}

\begin{proof}
Formula for $C_U(f,g)$ is a direct consequence of Theorem \ref{thm:C_N} and Theorem \ref{thm:TT2}. Since $U_f, U_g$ and $U_{f \circ g}$ correspond to unitary operators, $C_U(f,g)$ takes values in $S^1 \cong \{ z \in \mathbb{C}; |z|=1\}$ and the real part of its  logarithm vanishes. 
\end{proof}

\begin{rem}
Explicit expressions for the left and the right hand sides of equation \eqref{eq:C_N=TT} are of very different nature. The determinant on the left hand side uses Grunsky coefficients of 
the welding components $f_-$ and $g_-$. On the right hand side, the TT energy functional  is 
an integral of local expressions in terms of all welding components $f_\pm, g_\pm, (f \circ g)_\pm$. It would be interesting to find a direct proof of the surprising equality \eqref{eq:C_N=TT} between these expressions.
\end{rem}

\begin{rem}
The imaginary part ${\rm Im} \,\log(C_U(f,g))$  of the cocycle $C_U(f,g)$ is an additive real valued group 2-cocycle on ${\rm Diff}^+(S^1)$. Assuming that 
$$
H^2({\rm Diff}^+_{\rm hol}(S^1), \mathbb{R}) =H^2({\rm Diff}^+(S^1), \mathbb{R}) \cong \mathbb{R},
$$ 
this cocycle must be cohomologous to the Bott-Virasoro cocycle. It would be interesting to find an explicit expression for the coboundary which represents their difference.
\end{rem}


\begin{thebibliography}{99}

\bibitem{AS1}
A. Alekseev, S. L. Shatashvili,
{\em Path integral quantization of the coadjoint orbits of the Virasoro group and 2-d gravity},
Nucl. Phys. {\bf B 323} (1989) no. 3, 719-733.

  \bibitem{Alekseev_Shatashvili_char_orbits_DH} 
  A. Alekseev, S. L. Shatashvili, 
  {\em Characters, coadjoint orbits and Duistermaat-Heckman integrals}, 
  J. Geom. Phys. {\bf 170} (2021) 104386, 20pp.
  
  \bibitem{AS2}
  A. Alekseev, S. L. Shatashvili,
  {\em Coadjoint Orbits, Cocycles and gravitational Wess-Zumino},
  Rev. Math. Phys. {\bf 30} (2018) no. 6, 1840001, 15pp.
  
  
  \bibitem{Berezin} 
  F. A. Berezin, 
  {\em The method of second quantization},
  Pure and Applies Physics {\bf 24}, Academic Press, New York - London 1966.
  
  \bibitem {Kirillov} 
  A. A. Kirillov, 
  {\em The orbits of the group of diffeomorphisms of the circle, and local Lie superalgebras},
  Funk. Anal. Prilozh. {\bf 15} (1981) no. 2, 75-76.
  
  
  \bibitem{Kirillov_welding}
  A. A. Kirillov,
  {\em K\"ahler structure on $K$-orbits of a group of diffeomorphisms of the circle},
  Funk. Anal. Prilozh. {\bf 21} (1987) no. 2, 42-45.
  
  \bibitem{LP} 
  V. F. Lazutkin, T. F. Pankratova,
  {\em Normal forms and versal deformations for Hill?s equation},
 Funk. Anal. Prilozh. {\bf 9} (1975) no. 4, 41?48.

\bibitem{Polyakov}
G. L. Pimentel, A. M. Polyakov, G. M. Tarnopolsky,
{\em Vacuum Decay in CFT and the Riemann-Hilbert Problem},
Nucl. Phys. {\bf B 907} (2016) 617-632.

\bibitem{Pom}
C. Pomerenke,
Boundary behaviour of conformal maps, 
Springer-Verlag, Berlin, 1992.

\bibitem{Sullivan}
S. Nag, D. Sullivan.
{\em Teuchm\"uller theory and the universal period map via quantum calculus and the $H^{1/2}$ space on the cicle},
Osaka J. Math. {\bf 32} (1995) 1-34.



\bibitem{SSS} 
P. Saad, S. Shenker, D. Stanford,
{\em JT gravity as a matrix integral},
preprint arXiv:1903.11115.


\bibitem{Segal} 
G.\ Segal,
{\em Unitary representations of some infinite dimensional groups},
Commun. Math. Phys. {\bf 80}  (1981) no. 3, 301-342.
  

  \bibitem{Stanford_Witten} 
  D. Stanford,  E. Witten,
  {\em Fermionic localization of the Schwarzian theory},
  J. High Energy Phys. 2017.10 (2017) 1-28.
  
  \bibitem{TT} 
  L. A. Takhtajan, L.-P. Teo,
  {\em Weil-Petersson metric on the universal Teichm\"uller space},
  Mem. Amer. Math. Soc. {\bf 183} (2006) no. 861.
 
  \bibitem{W1} 
  E. Witten, 
  {\em Coadjoint Orbits of the Virasoro Group}
  Commun. Math.  Phys. {\bf 114} (1988) 1.

\end{thebibliography}
\end{document}